\documentclass[12pt]{iopart}
\usepackage{graphicx} 
\usepackage{dcolumn}
\usepackage{bm}
\usepackage{xcolor}
\usepackage{amssymb}
\usepackage[a4paper]{geometry}
\usepackage{subcaption}
\usepackage[utf8]{inputenc}
\usepackage[T1]{fontenc}
\usepackage{braket}

\renewcommand{\t}[1]{\mathrm{#1}}
\newcommand{\p}[0]{\mathrm{p}}
\renewcommand{\vec}[1]{\mathbf{#1}}
\newcommand{\Fermi}[0]{\epsilon_\t{F}}

\begin{document}

\title[Advancing our Understanding of Optoionic Effects]{Advancing our Understanding of Optoionic Effects\\ for the Design of Solar Batteries:\\ A Theoretical Perspective}

\author{Matteo Rinaldi$^1$, Matthias Kick$^1$,\\
Karsten Reuter$^1$, and Christian Carbogno$^1$}
\address{$^1$Fritz-Haber-Institut der Max-Planck-Gesellschaft, Faradayweg 4-6, 14195 Berlin, Germany}
\ead{carbogno@fhi-berlin.mpg.de}

\date{\today}

\begin{abstract}
Optoionics, a promising new field that aims at controlling ion dynamics using light, links photovoltaic power generation with electrochemical charge storage. This has the potential to drive and accelerate the energy revolution by utilizing materials that integrate the functionality of batteries and photovoltaic cells. Finding, optimizing, and customizing these materials is a complex task, though. Computational modeling can play a crucial role in guiding and speeding up these processes, particularly when the atomic mechanisms are not well understood. This does however require expertise in various areas, including advanced electronic-structure theory, machine learning, and multi-scale approaches. In this perspective, we shed light on the intricacies of modeling optoionic effects for solar battery materials. We first discuss the underlying physical and chemical mechanisms, as well as the computational tools that are available to date for describing these processes. Furthermore, we discuss the limits of these approaches and identify key challenges that need to be tackled to advance this field.
\end{abstract}

\maketitle

\section{Introduction}
\label{sec:Intro}
In close analogy to the optoelectronic effect, the term {\it optoionic effect}~\cite{Senocrate.2020} has been recently coined to describe scenarios in which light can be used to control the dynamics of ions and the associated macroscopic properties,~e.g.,~the ionic conductivity. This has for instance been demonstrated for methylammonium lead iodide in the seminal experimental work by Kim {\it et al.}, in which an enhancement of the ionic conductivity by two orders of magnitude was induced by illumination with above-band-gap energies~\cite{kim2018large}. At an atomistic level, this effect was explained by a change in ionic charge of the iodine atoms when Frenkel-type vacancy-interstitial pairs are created, which in turn are induced by the light-driven formation of excitons and their subsequent decay. As this example already highlights, the optoionic effect is typically mediated by electronic processes,~i.e.,~here by the electron-hole pair decay. Accordingly, it would in fact be more appropriate to refer to this kind of light-ion coupling as {\it opto-electro-ionic} effect.

Optoionic effects hence connect two typically unrelated fields: Optoelectronics, most notably its application in photovoltaics, and solid state ionics, a crucial field of study for energy storage and conversion. In turn, a unification of these fields promises unprecedented technologies at the interface between photovoltaics, photocatalysis, and electrochemical energy storage. Let us just mention a few example applications that could be enabled by this effect. For instance, optoionic effects could be exploited to probe and control the next generation of solid-state batteries, enabling faster and more efficient charging via illumination~\cite{lv2022photoelectrochemical}. In a very similar spirit, one could aim at using light for controlling (catalytic) chemical reactions, by enhancing or suppressing the flow of required ions and charges, as for instance realized in solar flow batteries~\cite{Jin.2018}. Last, but certainly not least, this technology holds great potential for sensing and diagnostic applications~\cite{Podjaski.2020}. 
\begin{figure}[t]
\centering
\includegraphics[width=0.55\textwidth]{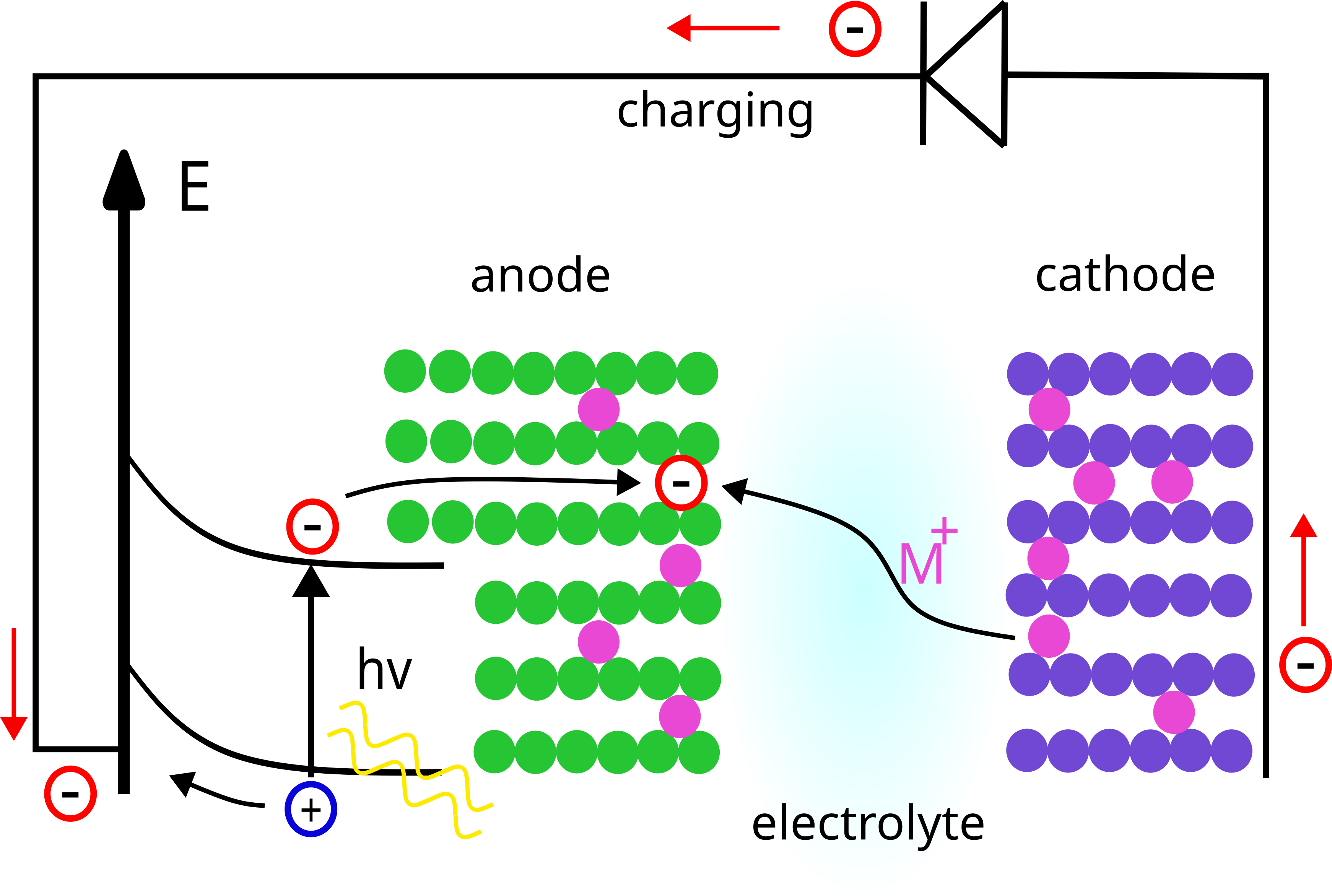}
\caption{Sketch of the working principle of a solar battery. Light generates excitons, which separate into two lattice-trapped, but mobile charges -- a hole and an electron polaron, respectively. In the shown example, the hole polarons recombine with electrons at the cathode through the external circuit, while the electron polarons interact with the intercalating metal ions.}
\label{fig:battery}
\end{figure}
One particular fascinating potential application of the optoionic effect are so-called ``solar batteries''. This term was recently coined to denote functional materials that encompass the functionality of both a photovoltaic and a long-term charge storage device within one compound~\cite{Gouder.2023n19}. In such systems, light is directly used to drive the intercalation of ions into the material, effectively resulting in the battery being charged (Figure \ref{fig:battery}). Although the basic idea was first demonstrated in the 1970ies~\cite{HODES.1976}, the underlying concepts were only more recently revived and several advanced device designs were proposed~\cite{lv2022photoelectrochemical,kandpal2023multifunctional}. Compared to a two-component solution featuring a separate photovoltaic and a charge-storage device, solar batteries offer the possibility of a higher level of integration. This could allow to eliminate losses associated to charge-extraction from the solar cell, wiring, and voltage or current mismatch. This is particularly appealing for decentralized applications,~e.g.,~in short-term solar energy buffers for alleviating network pressure. The same concept also enables to follow nature's example of photosynthesis and to temporally separate light and dark reactions in photocatalysis,~e.g.,~for time-delayed production of hydrogen from solar energy in the dark~\cite{Lau.2017,Wang.2024}. Finally, let us also note that this unification of multiple mechanisms within one material facilitates miniaturization and nanofabrication, as for instance demonstrated by light-driven microswimmers~\cite{Sridhar.2020}.

\begin{figure}[t]
\centering
\includegraphics[width=1.0\textwidth]{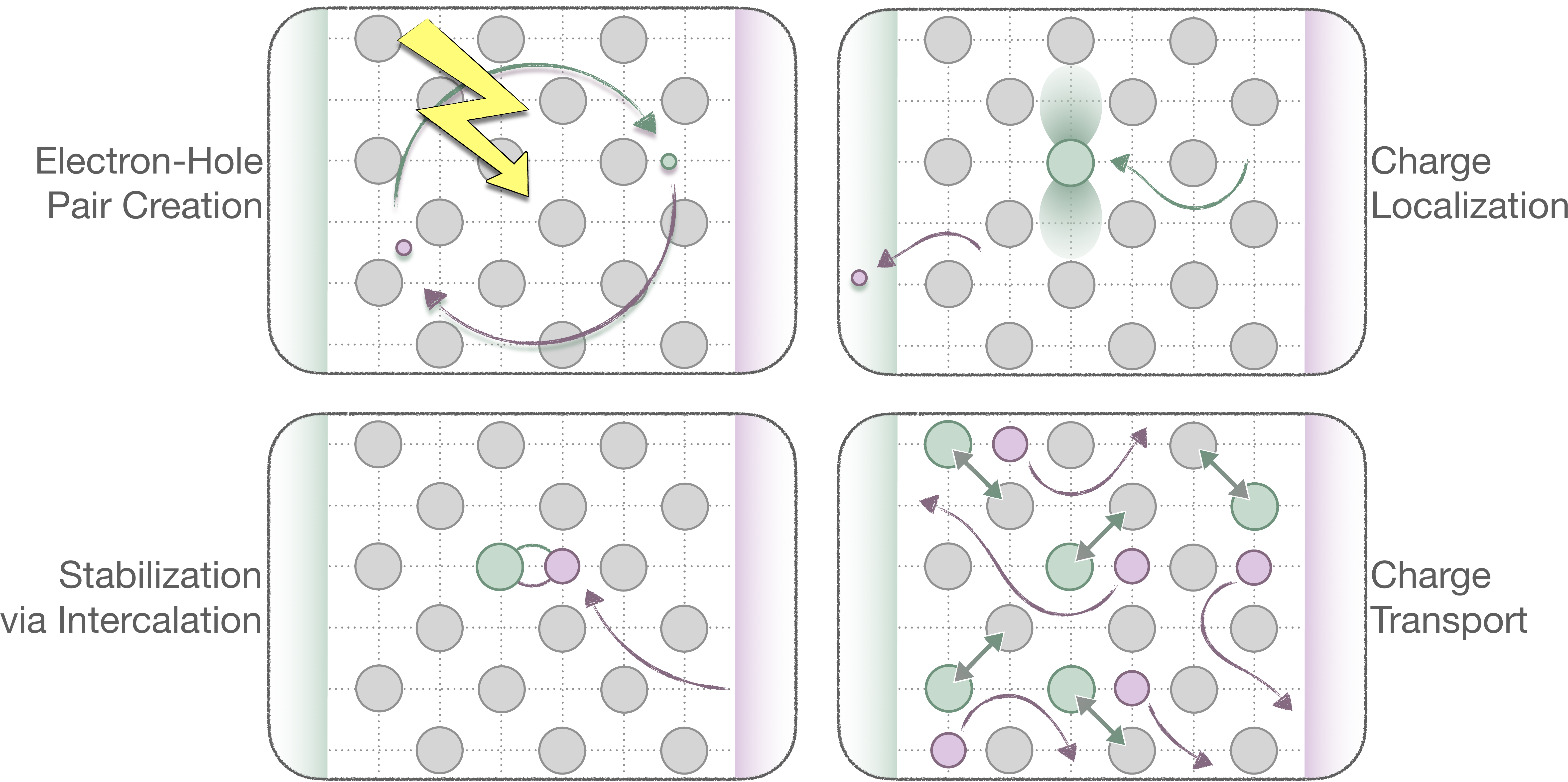}
\caption{Sketch showcasing the four fundamental physical processes, often subsumed under the term {\it optoionic} effect, that underlay the functionality of a solar battery material.}
\label{fig:sketch}
\end{figure}

\begin{figure}[t]
    \centering
    \includegraphics[width=0.55\textwidth]{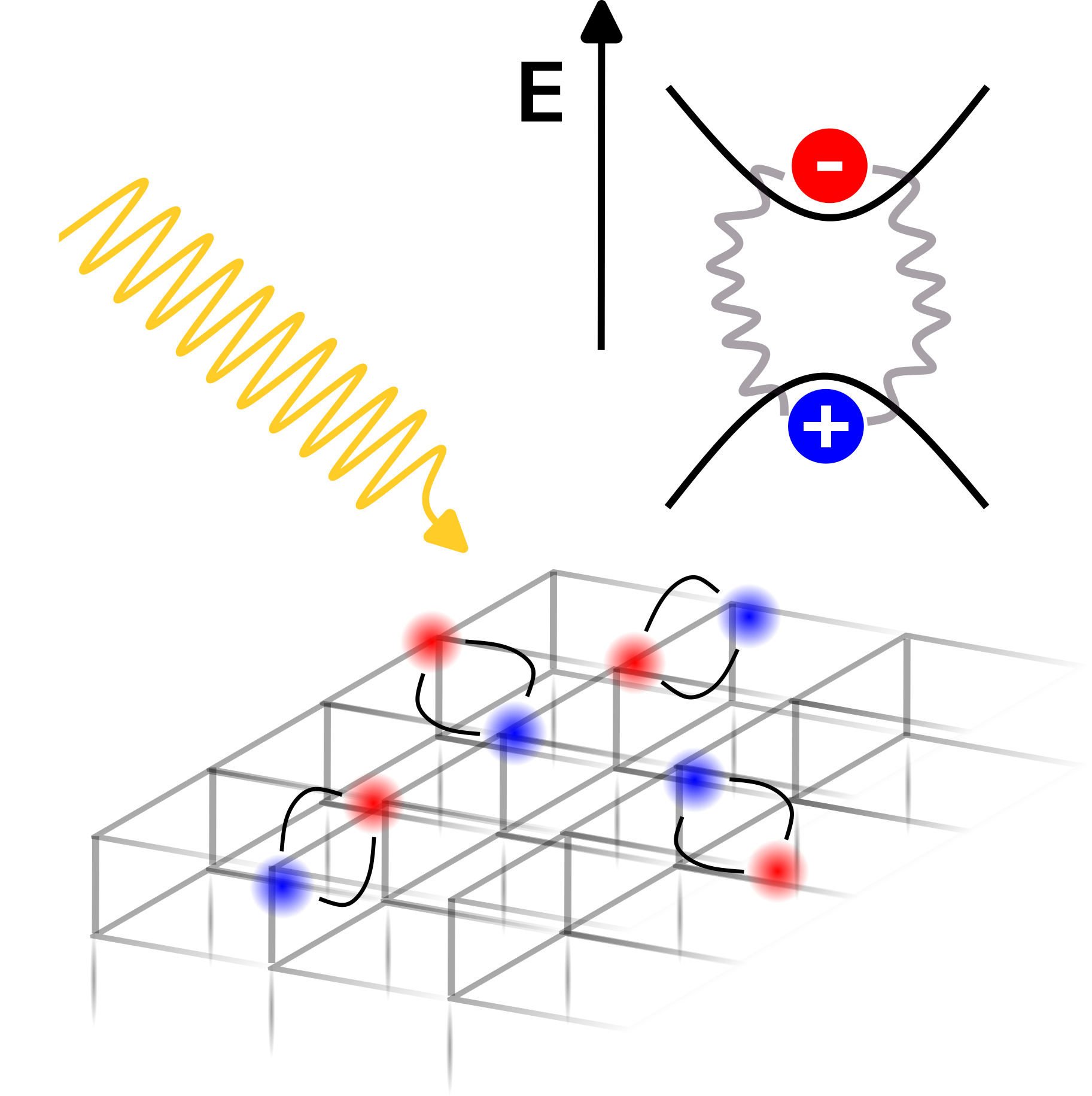}
    \caption{Incoming radiation generates electron-hole pairs (excitons) within the crystal lattice. The excited electron leaves behind a region of positive charge (hole). Electron and hole are bound together via Coulomb and exchange-correlation interactions. To model these interactions adequately one needs to go beyond standard ground-state electronic structure methods.}
    \label{fig:exciton}
\end{figure}

While this high level of integration is attractive for applications, it constitutes a severe hurdle for a theoretical and computational modeling, since several quite distinct, but coupled physico-chemical processes have to be addressed (in principle at the same time). These processes range from ultrafast electronic excitations at the atto- to femtoscale over atomistic dynamics at the nano- and microscale up to macroscopic effects driven by mass, charge, and energy flows. Accordingly, a wide range of expertise and computational techniques need to be leveraged for understanding and designing solar-battery materials (SoBaMs). This becomes evident when inspecting the individual steps, which are also sketched in Figure~\ref{fig:sketch}:
\\\noindent
{\bf Electron-Hole Pair Creation:} Light absorption can excite electrons to a higher energy state, leaving behind a hole in the originally occupied state (Figure \ref{fig:exciton}). This electron and hole attract each other via Coulomb interactions and can form a hydrogen-like quasi-particle, the so called {\it exciton}~\cite{doi:10.1126/sciadv.adi1323}. This electron and hole pair can further separate, recombine, or localize, see below. To understand these processes, it is hence required to model electronically excited states and their dynamics. The involved timescales are typically well-below a picosecond; still, the exciton creation and subsequent dynamics sensitively depend on the chemical environment,~e.g.,~the band gap and the localized occupied or unoccupied defect and polaron states within the gap.
\\
\noindent{\bf Charge Localization:} To enable charge storage, it is essential that the electron-hole pair separates and that (at least one of the two) excess charges are retained within the material. In SoBaMs this retention is achieved through a localization of the excited-state wave function through interaction with the nuclei and a concomitant distortion of the local geometric structure. When this results in localization on a single atomic site, this process can be understood as changing the chemical oxidation state.
More generally, both such highly localized and more delocalized excess charges can be described in terms of polaron formation.
Since such localized charges are largely stabilized by geometric distortions, this occurs at typical lattice-dynamics timescale, roughly in the picosecond regime. Again, the localization process depends sensitively on the local and global chemical environment, given that excess charges interact via long-ranged Coulomb forces. Concurrently, the localization is in competition with ultrafast processes like charge recombination.
\\\noindent
{\bf Charge Stabilization via Intercalation:} The repulsive interaction between localized excess charges prevents the realization of high energy densities as would be required in applications like batteries. This hurdle can be circumvented by compensating the excess charges through uptake of ionic species of opposite charge. In layered materials frequently employed in batteries, such an incorporation of mobile ions is called intercalation. The involved chemical processes occur at typical kinetic regimes, ranging from pico- to nanoseconds. Again, they depend sensitively on the available excess charges, on the overall state-of-charge of the battery material,~i.e.,~the global chemical environment defined by the degree of intercalation, and on the availability of intercalating elements, see below.
\\\noindent
{\bf Charge and Mass Transport:} Batteries do not only require high energy densities to be technologically relevant, but also a rapid charge and discharge rate, so to enable high-power applications. Along these lines, charge and discharge are required to be highly reversible to maximize the device's lifetime. To the utmost extent, these boundary conditions are determined by non-equilibrium thermodynamics,~i.e.,~by the transport coefficients. These describe how fast charge, mass, and heat can be transported to where they need to be for enabling charge localization and stabilization. Being thermodynamic processes at the macroscale, the involved timescales can range up to microseconds and beyond, even though the actual microscopic dynamics occurs much faster.
\\

One practical, well-studied example for such an anode solar-batter material~(SoBaM) are 2D carbon nitrides in solution, in particular potassium polyheptazine imide (K-PHI)~\cite{Podjaski.2018,Podjaski.2020,Gouder.2023e0v}. In this compound, the photogenerated holes are extracted via the electrolyte by a redox shuttle, whereas the photogenerated electrons localize on the heptazine units. These excess charges are further stabilized on macroscopic time scales,~i.e.,~over several hours, by an influx of K$^{+}$ cations from the electrolyte through the structural pores. Similar optoionic mechanisms have, for instance, been reported for solvated 2D NbWO$_{6}$~\cite{Wang.2024}, in which intercalated Li$^+$/H$^+$ ions stabilize polarons on the tungsten sublattice, for 2D perovskites, in which the photogenerated holes drive Li$^+$ out of the perovskite and induce photocharging~\cite{Ahmad.2018}, and for MoO$_{3}$, in which the photogenerated electrons are stored and stabilized by the intercalation of Na$^{+}$~\cite{Lou.2017}. 

As the above examples highlight, a practical realization of solar batteries is most easily achieved in compounds that exhibit structural and stoichiometric complexity. Naturally, this enables virtually endless possibilities for optimization. Among others, appealing optimization targets include increased light-conversion efficiency, improved charge trapping, and increased electronic and ionic charge transport. In this regard, computational approaches lend themselves to avoid laborious and costly experiments and can hence be pivotal in steering and guiding materials' space exploration for solar-battery discovery. Still, this is far from trivial due to the inherent complexity of candidate materials and due to the richness in phenomena that need to be modeled. Even though the above described steps occur on time scales that vary by orders of magnitude, they can often not be considered independently. Rather, these processes occur continuously and concurrently in an actual solar battery and severely influence each other.

In this perspective, we aim at providing an overview on the theoretical and data-driven methods that can aid and guide the design and optimization of solar battery materials~(SoBaMs). To this end, we will first discuss the individual processes involved in the optoionic effect in more detail in Sec.~\ref{SoA}. Given the variety of time and length scales as well as of expertise that needs to be covered, we here also provide a concise overview over the different computational approaches that are available today for studying the individual processes. In this spirit, Sec.~\ref{Chall} will then focus on when, how, and why these techniques often reach their limits for the modeling of SoBaMs. In particular, we will focus on those hurdles that can and should be overcome to facilitate an {\it in silico} solar battery optimization and which approximations can help in this endeavor. Eventually, Sec,~\ref{Concl} summarizes these concepts and explores potential applications of these developments beyond the realm of SoBaMs.

\section{State-of-the-Art Theoretical Approaches}
\label{SoA}

\subsection{Electron-Hole Pair Creation and Decay} 
\label{sec:EHP}
To simulate and to understand electron-hole pair creation and their dynamics, a static ground-state approach is obviously inadequate and it is hence necessary to explicitly target excited electronic states. This is particularly challenging as one has to consider the interaction of the electrons with the electro-magnetic field of the photon. How a system of electrons will react to the field of a photon with a certain frequency can be obtained by explicitly calculating the density response function of the system~\cite{doi:10.1021/cr0505627}. The poles of the response function are then exactly the points where electron motion will be in resonance with the photon frequency and absorption will occur~\cite{annurev.physchem.55.091602.094449}. This also naturally explains why only photons with specific energies can be absorbed. 

In general, two different aspects of these excitations are of particular interest for modeling SoBaMs. First, it is necessary to compute the energetic spectrum of excited states including their oscillator strengths. This allows to address the formation of electron-hole pairs and to predict at which wavelengths the system will absorb (and emit) light with which yield in electron-hole pairs. Second, it is essential to simulate how these states evolve after such an excitation event by tracking population changes and structural reconfigurations that occur over time. This allows to pinpoint the mechanisms underlying charge separation and to determine how efficiently localized charge storage is realized via photophysical and photochemical processes. 

As discussed later in Sec.~\ref{Chall}, modeling SoBaMs typically requires to target large system sizes. However, strategies such as the algebraic diagrammatic construction technique~(ADC)~\cite{C7CP07849H}, methods based on many-body perturbation theory such as $GW$ and the Bethe-Salpeter Equation (BSE)~\cite{PhysRevLett.128.016801}, and the equation-of-motion coupled-cluster approach~(EOM-CC)~\cite{doi:10.1021/acs.jctc.0c00639,10.1063/5.0004865,10.1063/5.0099192} typically exhibit a very unfavorable, at least quartic scaling with system size. In this light, time-dependent density-functional theory~(TDDFT)~\cite{progress_tddft} with cubic or even lower scaling appears to be the most promising technique for the description of excited states in SoBaMs. For the exact same reasons TDDFT is the typical method of choice for modeling excitons in extended solid-state systems,~e.g.,~perovskite solar cells~\cite{doi:10.1021/acs.jpclett.9b00641}. Accordingly, we focus on TDDFT-based methods for the description of excited states in this perspective as well.

From a bird's eye view, two different formulations and applications of TDDFT can be discerned~\cite{doi:10.1021/ct500763y}. In linear-response TDDFT (LR-TDDFT), the first order response of the density to a time-dependent perturbation is determined. Excitation energies and oscillator strengths are then obtained by recasting the equations into a non-hermitian eigenvalue problem, which is subsequently solved in frequency space~\cite{progress_tddft}. The obtained eigenvalues correspond directly to the system's excitation energies, while the eigenvectors are used to calculate transition dipole moments and oscillator strengths. Additionally, they help determine hole and electron localization, as well as quantum yields.~\cite{D2RA06880J,C3CP51514A} Note that LR-TDDFT also allows to calculate excited state forces useful for excited state geometry optimization or dynamics.~\cite{C3CP51514A}

In contrast, real-time TDDFT~(RT-TDDFT)~\cite{doi:10.1021/ct500763y} explicitly propagates the electronic states in time, whereby a time-dependent external potential can be explicitly included. Formally, this is closely related to solving a quantum Liouville equation~\cite{C5CP03712C} and hence naturally allows to explore the dynamics of excited states. Additionally, this allows to incorporate the effects of nuclear motion and/or structural reorganization. In the simplest case, the nuclear motion is incorporated in a classical sense meaning that the nuclei move according to classical Newtonian mechanics, while the electrons evolve quantum mechanically (Ehrenfest dynamics)~\cite{10.1063/1.1856460,10.1063/1.2008258}. Besides providing the temporal evolution of the wave function, RT-TDDFT also allows to determine the excitation spectrum by inspecting the Fourier-transformed dipole-dipole autocorrelation. This is particularly useful for studying nonlinear effects, since RT-TDDFT is a non-perturbative approach that also accounts for effects beyond linear response. For instance, this fact has been successfully exploited for studying plasmon excitations~\cite{doi:10.1021/acsnano.8b08703,doi:10.1021/acsnanoscienceau.2c00061,doi:10.1021/acs.jpcc.7b04451}, photo-induced charge transfer at heterojunctions~\cite{Rozzi_2018,doi:10.1126/science.1249771}, the density response to ultrafast, intense, or short-pulse lasers~\cite{https://doi.org/10.1002/qua.25096}, and non-equilibrium electron dynamics, see the excellent review by Xu {\it et al.}~\cite{kanai_perspective}.

\subsection{Charge Localization}
\label{sec:localization}
In their ground state, excess charges usually form delocalized states at the band edges in semiconducting or insulating solids. Excess holes are hence found at the valence band maximum~(VBM), while electrons are found at the conduction band minimum~(CBM) of the respective ``uncharged'' crystal. Localization of these charges can only be achieved through an additional driving field that can, for instance, stem from impurities (then resulting in a charged defect) or from lattice distortions (then resulting in a polaron). In both cases, an additional electronic state is formed within the band gap with negligible dispersion reflecting the localized real-space character.

Over the last decades, a theoretical framework has been developed to compute, understand, and describe defect formation and stability in the dilute limit,~i.e.,~for isolated defects~\cite{Freysoldt.2014}. This allows to rationalize and predict the properties of semiconductors for electronic and optoelectronic applications, which can be significantly altered by defects, even when only present in particle-per-million concentrations.
 
One key quantity for assessing the stability of a specific defect is its formation energy
\begin{equation}
E^\t{f}\left[ X^{Q} \right] = E_\t{tot}\left[ X^{Q} \right] - E_\t{tot}\left[ \t{bulk} \right] - \sum_{i}n_{i}\mu_{i} + Q\Fermi \;. 
\label{eq:ef}
\end{equation}
Here, $E_\t{tot}\left[ X^{Q} \right]$ and $E_\t{tot}\left[ \t{bulk} \right]$ correspond to the total energy of the system with and without defect, while $n_{i}$ and $\mu_{i}$ denote the number of atoms of type $i$ that have been added or removed in the defected system and their chemical potentials. Accordingly, this term balances changes in stoichiometry. Similarly, the next term balances excess charges with $Q$ being the charge of the defect and $\Fermi$ being the Fermi level that serves as charge reservoir. 

To address vanishingly small concentrations that are typically of interest in semiconductor physics, (\ref{eq:ef}) needs to be evaluated in the dilute limit, so to obtain the formation energy of a single defect. To this end, extended supercells are necessary so that the interaction between the defect and its artificial periodic images becomes negligible. Especially in the case of charged defects interacting via long-ranged electrostatic interactions, this often requires impractically large supercells spanning thousands of atoms. To circumvent this issue, extrapolation techniques~\cite{Leslie.2000,Makov.1995} or electrostatic correction schemes~\cite{freysoldt_correction,Kumagai.2014} can be employed to obtain correct values for the bulk limit within finite supercells. Still, even in this case, the supercells have to be large enough to ensure that there is no overlap between the wave function of the defect and the one of its periodic images, so that a truly localized electronic state is realized.

The above-described formalism allows to investigate the defect-formation energy~$E^\t{f}\left[ X^{Q} \right]$ as a function of the Fermi-energy~$\Fermi$,~i.e.,~in a grand canonical picture~\cite{Freysoldt.2014}. Furthermore, it is possible to incorporate thermodynamic effects at finite temperatures. To this end, electronic, vibrational, and configurational entropic effects can be accounted for, yielding temperature- and pressure-dependent Gibbs' free energies of formation~\cite{grabowski2009ab,glensk2014breakdown}. 

As already mentioned above, a stoichiometric defect is often not even needed to localize excess charges. In solids featuring strong ionic dielectric screening, lattice distortions alone can be sufficient to generate a localization field, resulting in a self-trapped charge or polaron. In principle, such polarons can be computationally tackled using the exact same framework used for defects, including (\ref{eq:ef}), in which $n_i=0$ is set to zero due to the absence of stoichiometric changes and $\Fermi$ is chosen as the respective band edge,~i.e.,~VBM for holes and CBM for electrons. The resulting polaron formation energy can then also be recast as:
\begin{equation}\label{eq:ef_falt}
E^\t{f}\left[X^{Q}\right] = Q\left(\Fermi - \epsilon_{p}\right) + \left[E_\t{tot}\left[X^{Q=0}\right]-E_\t{tot}\left[\t{bulk}\right]\right],
\end{equation}
where $E_\t{tot}\left[X^{Q=0}\right]$ is the total energy in the neutral charged state (i.e. without polaron) and $\epsilon_{p}$ is the polaron energy level. In this expression the first term accounts for electronic localization, while the second relates to the lattice distortions induced by the polaron. 

\begin{figure}
    \centering
    \includegraphics[width=0.55\textwidth]{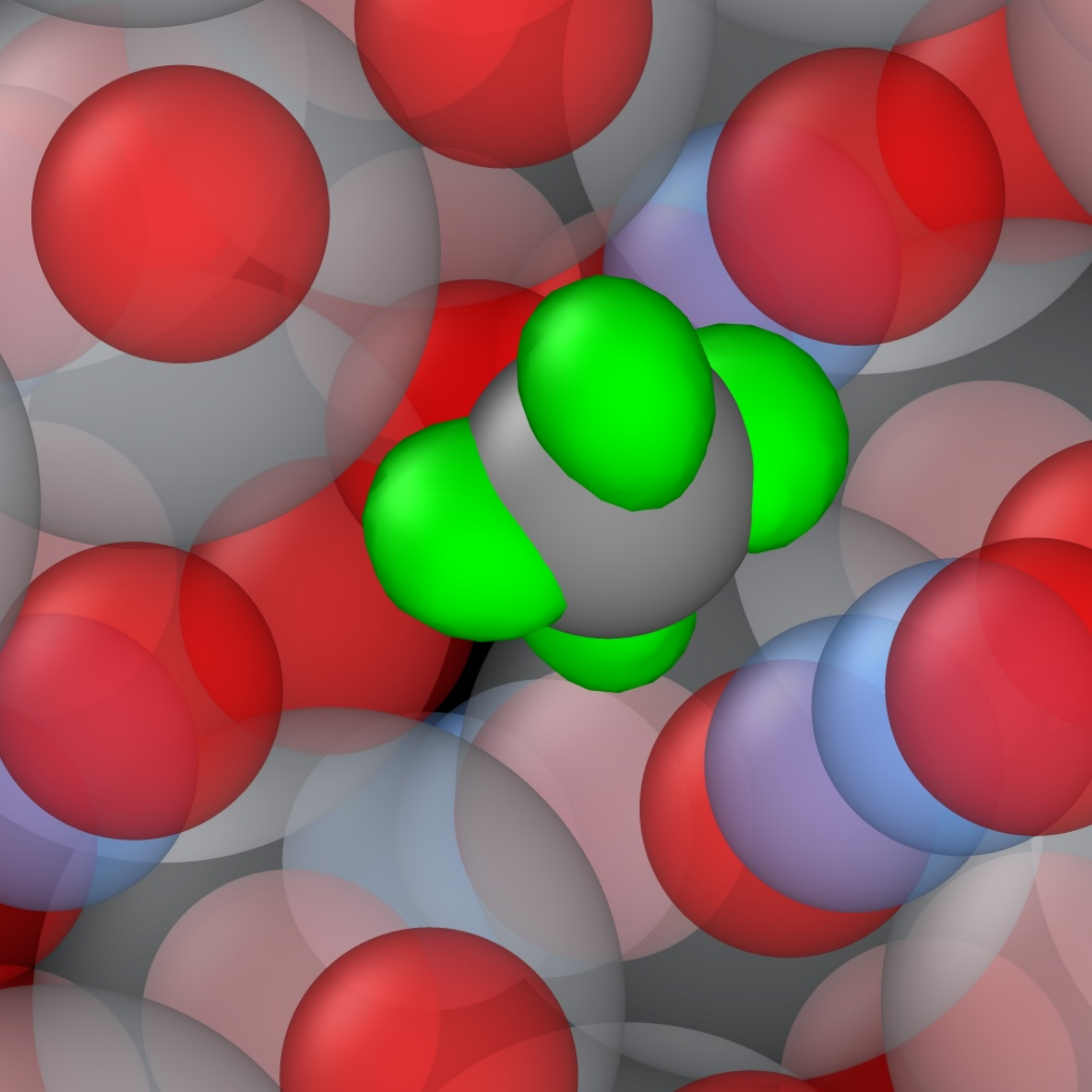}
    \caption{Isodensity surface~(10\% of maximum, light green) of an electron polaron state in Li$_{4}$Ti$_{5}$O$_{12}$ with Li shown in azure, Ti in grey, and O in red. The excess charge localizes on one Ti atom, highlighted in dark grey, and one hence obtains a small polaron.}
    \label{fig:polaron_lto}
\end{figure}

In practice, however, evaluation of (\ref{eq:ef_falt}) is computationally not always straightforward. Due to the fact that the localization field is not generated by a localized defect, but by potentially much more delocalized lattice distortions, polarons also come in much larger sizes. While so called ``small'' polarons are indeed as localized as charged defects and can be treated as such (see Figure~\ref{fig:polaron_lto} for a representative example), ``large'' polarons can span several hundreds if not thousands of atoms. Obviously, targeting supercells that are large enough to capture such large polarons,~i.e.,~that are extended enough so that there is no overlap between the wave function of the polaron and the one of its periodic images, is numerically quite costly. This is aggravated by the fact that it is \textit{a priori} largely unknown if and with which size a polaron can form at all. 

To overcome this computational issue, it is possible to approximate the problem and to revert to {many-body} perturbation theory, as proposed by Sio {\it et al.}~\cite{Sio.20190b,Sio.2019}. By this means, the real-space supercell problem can be recast into the solution of a non-linear eigenvalue equation in reciprocal space, hence avoiding the explicit usage of supercells. To this end, the total energy of the system including an excess charge is recast as
\begin{eqnarray}
E_\t{tot}^\p & \approx & E^\t{eq} + \frac{1}{2}\sum_{I,J} \Phi_{IJ}^\t{eq} \Delta\vec{R}_I \Delta\vec{R}_J \nonumber \\
                      && + \braket{\psi_\p | {H}^\t{eq}| \psi_\p} + \sum_I \braket{ \psi_\p| \frac{\partial V^{\p}}{\partial \vec{R}_I}| \psi_\p} \Delta\vec{R}_I  \;.
\label{eq:sio_start}
\end{eqnarray}
Here, the first pair of terms corresponds to the {\it harmonic} approximation for the neutral solid without excess charge,~i.e.,~a truncated second-order Taylor expansion of its potential-energy surface valid in the limit of small displacements~$\Delta\vec{R}_{I}=\vec{R}_{I}-\vec{R}_{I}^\t{eq}$ from the equilibrium position. Accordingly, $E^\t{eq}$ denotes the total energy at the equilibrium position without any excess charge and $\Phi_{IJ}^\t{eq}$ the system's Hessian. Hereby, $I$ and~$J$ are generalized atomic indices that include the respective Cartesian components and that run over all atoms in the infinite solid. Similarly, the second pair of terms describes the energetic changes upon addition of an excess charge with wave function~$\psi_\p$,~i.e.,~the sought-after polaron wave function. As in (\ref{eq:ef_falt}), this reflects the electronic reorganization energy up to linear order and hence includes the derivative of the potential acting on the electronic subsystem~$V^{\p}$ with respect to nuclear displacements upon addition of the excess charge. 

In the next step, the problem is recast in reciprocal-space, where $\vec{k}$ and $\vec{q}$ are associated to electronic and nuclear degrees of freedom. Accordingly, the wave function for the excess charge $\psi_\p \sim \sum_{n\vec{k}}A_{n\vec{k}}\psi_{n\vec{k}}^\t{eq}$ is expanded in terms of the equilibrium states of the neutral system~$\psi_{n\vec{k}}^\t{eq}$ and the displacements $\Delta \vec{R}_I \sim \sum_{\vec{q}\nu}B_{\vec{q}\nu}\vec{e}_{I,\vec{q}\nu}$ in terms of the phonon eigenvectors~$\vec{e}_{\vec{q}\nu}$  with phonon frequencies~$\omega_{s\vec{q}}$ obtained from the diagonalization of the harmonic Hamiltonian introduced above. By variationally minimizing (\ref{eq:sio_start}), one eventually obtains a secular equation for an isolated, non-self interacting\footnote{Since one deals with a one-electron problem in (\ref{eq:sio_start}), the spurious self-interaction terms of the polaron with itself and with its periodic images can and have been eliminated.} polaron, which can be solved self-consistently to determine the expansion coefficients~$A_{n\vec{k}}$ and $B_{\vec{q}\nu}$. In this approximation, the polaron formation energy can be expressed as
\begin{equation}\label{eq:ef_sio}
E^\t{f}\left[X^{p}\right] = \sum_{n\vec{k}}\left|A_{n\vec{k}}\right|^{2}\left(\varepsilon_{n\vec{k}}-\Fermi\right) - \sum_{\vec{q}\nu}\left|B_{\vec{q}\nu}\right|^{2}\hbar\omega_{\vec{q}\nu},
\end{equation}
where the first and second contribution are purely electronic and vibrational, respectively. In other words, this then corresponds to a recasting of (\ref{eq:ef_falt}) in a many-body perturbation theory framework. Without going into further refinements of this ansatz~\cite{lee2021facile,Lafuente-Bartolome.2022,lafuente2022_prb,luo2022comparison}, let us just stress that these kind of approaches work best for highly-ordered systems with few atoms per unit cell, for which a representation in terms of well-defined states in reciprocal space is easily possible.

\subsection{Charge Stabilization}
\label{sec:stabilization}
As discussed in the introduction, a long-term stabilization of the polaronic charge is typically achieved via counterion intercalation. While modeling such processes is in principle straightforward, the sheer amount of structural and compositional modifications and the immense amount of atomistic pathways leading to such configurations rapidly leads to a combinatorial explosion, even in the simplest systems. Efficiently exploring such high-dimensional spaces via (accelerated) molecular dynamics and/or Monte Carlo approaches is a key research subject in computational statistical mechanics and several excellent textbooks are devoted to this topic~\cite{Frenkel.2002,Tuckerman.2023}. Still, even the most efficient electronic-structure theory techniques are too costly for a the brute-force evaluation of such sampling algorithms. Accordingly, the costly $\textit{ab initio}$ calculations have increasingly been supplanted in recent years by machine-learned interatomic potentials (MLIPs), which typically exhibit a linear scaling behavior. These MLIPs enable access to longer time scales and larger system sizes while offering reduced computational cost and greater accuracy compared to traditional empirical force fields, such as embedded atom and bond order potentials. The emergence of MLIPs was essentially favored by their much more flexible functional form and by the development of system representations in form of local atomic environments that facilitate the learning either from large databases of $\textit{ab initio}$ calculations or through active learning where new {\em ab initio} calculations are queried on demand. In particular, approaches such as neural networks based on Behler-Parrinello symmetry functions~\cite{behler2007generalized}, Gaussian approximation potentials (GAP)~\cite{bartok2010gaussian}, Moment Tensor Potentials (MPT)~\cite{shapeev2016moment}, Atomic Cluster Expansion (ACE)~\cite{drautz2019atomic} and more recently frameworks based on equivariant neural networks such as NequIP~\cite{Batzner2022}, MACE~\cite{batatia2022mace}, grACE~\cite{bochkarev2024graph}, and SO3KRATES~\cite{Frank2024} have been successfully applied to a wide range of systems with increasing complexity. 
As it is the case in traditional,~i.e.,~non-solar batteries,~modeling the intercalation processes for cathode, anode, and electrolyte as well as for the respective interfaces with MLIPs can be challenging and time-consuming, as this requires many training structures with diverse chemical and geometrical diversity to capture all the relevant interactions and processes~\cite{deringer2020modelling}. Foundation models~\cite{takamoto2022towards,deng2023chgnet,merchant2023scaling,batatia2023foundation,zhang2024dpa,yang2024mattersim},  which are trained on large datasets that encompass all kind of structures, stoichiometries, and processes, can help mitigate these challenges. Although such general-purpose models do presently not reach the high precision of a model that is specifically trained for one individual problem, they are typically accurate enough to boot-strap the learning process. Obviously, refining a foundation model for a specific system of interest is computationally less involved than starting from scratch. In this spirit, foundation models can be employed to rapidly cover the thermodynamic phase space explored during dynamics; selected, uncorrelated configurations out of this simulations can then be refined from first principles and serve as training set for the refinement.

\subsection{Charge Transport}
\label{sec:transport}
\begin{figure}[t]
\centering
\includegraphics[width=0.8\columnwidth]{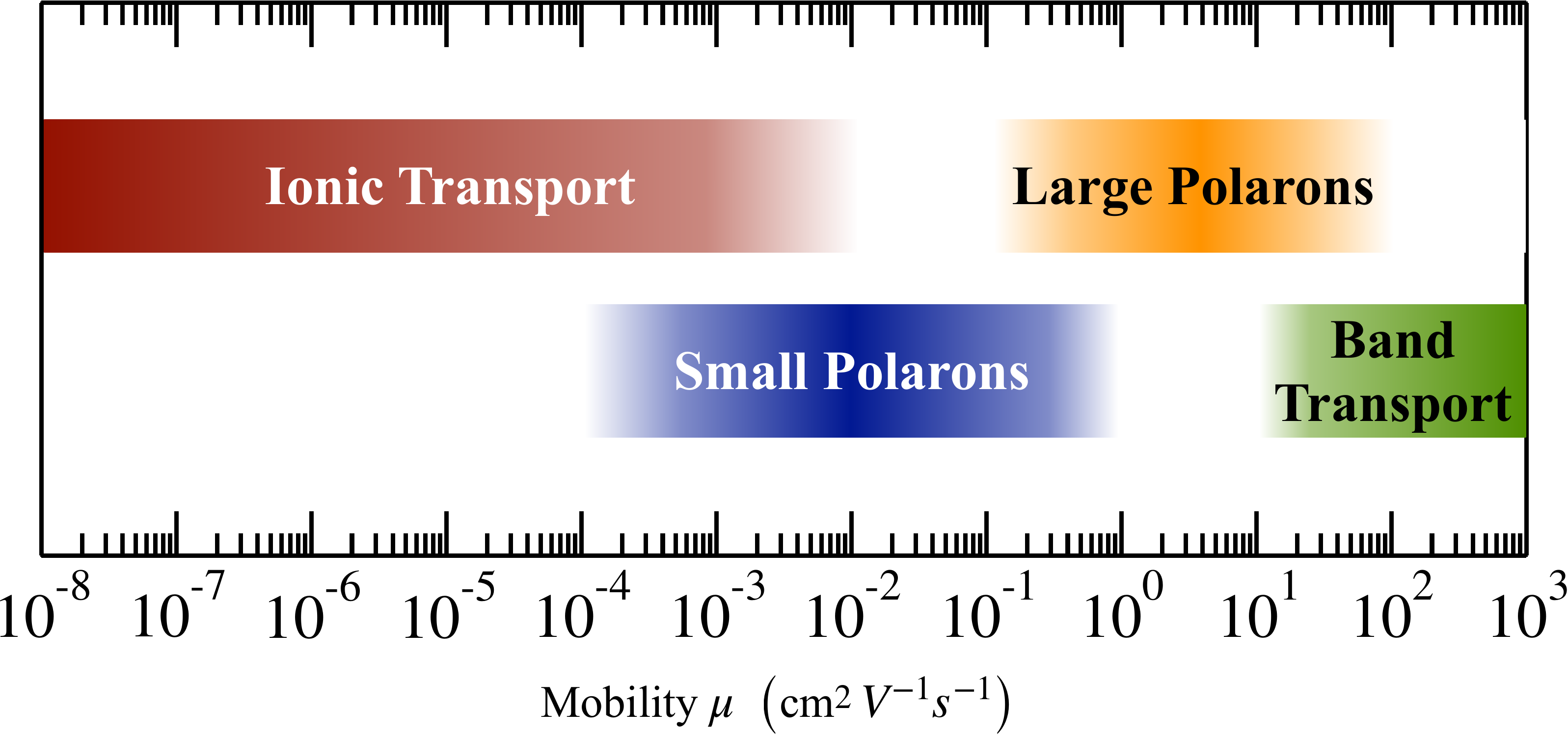}
\caption{Typical mobility ranges of ionic, polaronic, and band conduction in solid phases~\cite{landolt1988numerical,Franchini.2021,chang2022intermediate,oberhofer2017charge,ponce2020first}.}
\label{fig:regimes}
\end{figure}
At the macroscopic level, charge transport is described by Ohm's law~$\vec{J}=\sigma\nabla U$, in which the conductivity~$\sigma$ reflects the proportionality between the charge flux~$\vec{J}$ and the potential gradient~$\nabla U$. This equation reflects that conduction is a diffusive process driven by thermodynamic fluctuations. One could, in principle, assess charge transport using slight variations of the dynamical techniques discussed in the previous subsections,~e.g.,~time-dependent electronic-structure theory and/or methods based on molecular-dynamics. In practice, however, reaching the time and length scales required to capture charge-transport becomes computationally prohibitive and additional approximations have to be taken. To this end, it is important to first clarify which type of conduction --electronic, polaronic, or ionic-- one aims to address. In this context, it also common to split up the conductivity~$\sigma=n\mu$ into two different contributions. The charge-carrier density~$n$, which describes how many (quasi-)particles are actually contributing to conduction, and the mobility~$\mu$, which captures how effective this contribution is. Since $n$ and $\mu$ behave quite different for different transport regimes, this distinction is particularly important for optoionic materials, in which different type of (quasi-)particles, including electrons, polarons, and ions might be involved in charge transport. 

\begin{figure}[t]
    \centering
    \includegraphics[width=1.0\columnwidth]{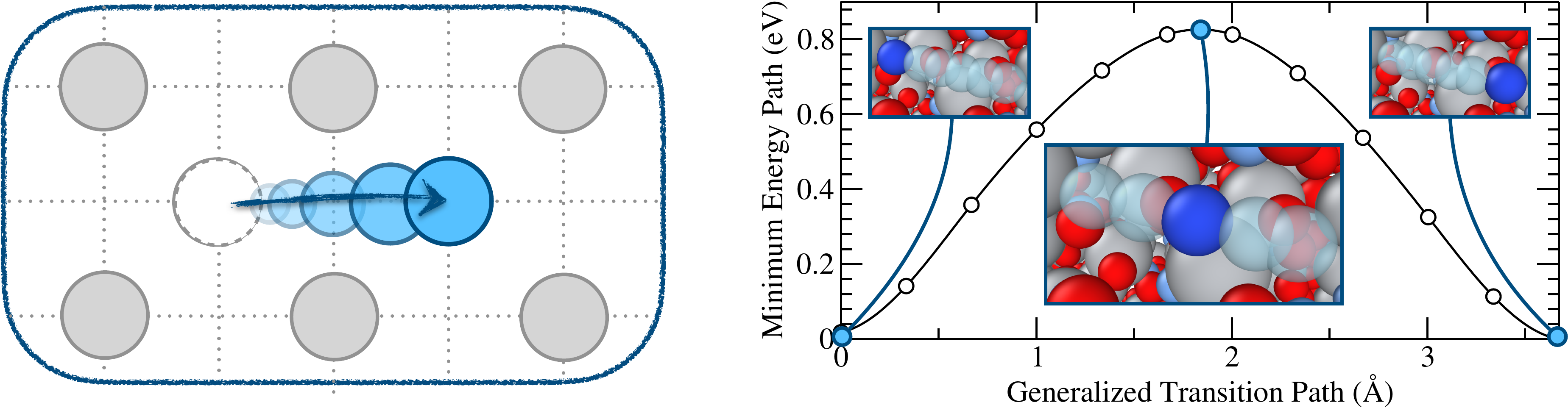}
    \caption{Ionic transport at an atomistic level: While the left sketch exemplifies the underlying dynamics, the right plot shows a representative minimum-energy path for lithium viz.~vacancy diffusion in Li$_{4}$Ti$_{5}$O$_{12}$. The inlets showcase the initial, transition, and final state using the same color code of Figure~\ref{fig:polaron_lto}. The motion and the exact location of the diffusing Li atom is highlighted in (semi-transparent) blue. At variance with {\it polaronic} transport, cf.~Figure~\ref{fig:transport_pol},  mass and charge transport go hand in hand in ionic transport.}
    \label{fig:transport_ion}
\end{figure}
For electronic or band-like transport, it is common to describe the charge-carrier dynamics in terms of independent quasi-particle states obtained in equilibrium~\cite{ponce2020first}. In semiconductors and insulators, one in turn obtains an exponential dependence of the charge-carrier density~$n$ on temperature and on the smallest gap between occupied and unoccupied electronic states. The latter corresponds to the fundamental band gap in the intrinsic limit, but can be sensitively reduced by shallow defects close to the band edge under extrinsic doping. The mobility~$\mu$ typically exhibits a $1/T^\alpha$ dependence and is limited by scattering processes: electron-electron, electron-defect, and electron-phonon scattering. The latter is typically dominant in intrinsic and modestly doped materials and can be modelled using the exact same many-body perturbation theory approximations introduced in Sec.~\ref{sec:localization}. To this end, the nuclear dynamics is modeled in terms of harmonic phonons and the electronic response using electron-phonon coupling elements~\cite{bernardi2016first,ponce2020first}; the resulting scattering cross-section can then be used to model this band-like transport using the Boltzmann and/or Wigner transport equation. The prior captures coherent transport in the semi-classical particle picture, whereas the latter describes incoherent wave-like transport, which typically dominates in more disordered systems. In other words, one assumes that charge transport is largely determined by the properties of the charges in the equilibrium system and that the nuclear motion can be treated as a minor perturbation that limits conduction. Note that there is no nuclear mass transport involved in this kind of approaches, since on thermodynamic average, all nuclei remain at their respective lattice sites and only the electrons or holes contribute to conduction. Note that also the influence of large and intermediate polarons on band transport can be incorporated in this formalism by extending many-body perturbation theory using a cumulant technique to capture higher-order electron-phonon interactions~\cite{Bernardi2019,chang2022intermediate}. Similarly, such higher-order electron-phonon couplings as well as anharmonic effects can be accounted for by employing the Kubo-Greenwood formalism~\cite{Kubo.1957jeb,Greenwood.1958,Holst.2011,Quan.2024}. 

Conversely, ionic transport does not require to consider electronic degrees of freedom independently, since charge can only move in conjunction with the respective ions, as shown in Figure~\ref{fig:transport_ion}. Accordingly, mass transport via nuclear motion needs to be explicitly taken into account and the perturbative limit of small nuclear displacements does no longer apply. In this regime, the number of charge carriers is either dictated during synthesis or, in the case of intercalation compounds, by the state-of-charge~(see Sec.~\ref{sec:stabilization}) and thermodynamic effects have only a minor influence on the charge-carrier density. Also, the mobility~$\mu$ increases exponentially with temperatures, since mass transport is an activated process that requires an energetic barrier to be overcome and hence follows Arrhenius-type relations. Typical computational approaches thus require to monitor the equilibrium dynamics in such compounds over large time scales by either using molecular-dynamics or kinetic Monte Carlo approaches. The mobility can then be extracted either by monitoring the mean squared displacement using Einstein-type relations or by monitoring the charge flux using Kubo-type relations. Let us emphasize that the respective ionic conductivity is not necessarily equal to the sum of diffusivities of the individual ionic species, since correlations between them can play an important role~\cite{he2017origin,xu2012one,jalem2013concerted,meier2014solid,burbano2016sparse,catti2011short,lang2015lithium,kubisiak2020estimates}. In passing, let us also note that a solid quantum-mechanical and formal foundation for the intuitive picture of ionic transport,~i.e.,~the one of individual atoms carrying a specific charge, was only established very recently from first principles~\cite{Grasselli.2019}.

\begin{figure}[t]
    \centering
    \includegraphics[width=1.0\columnwidth]{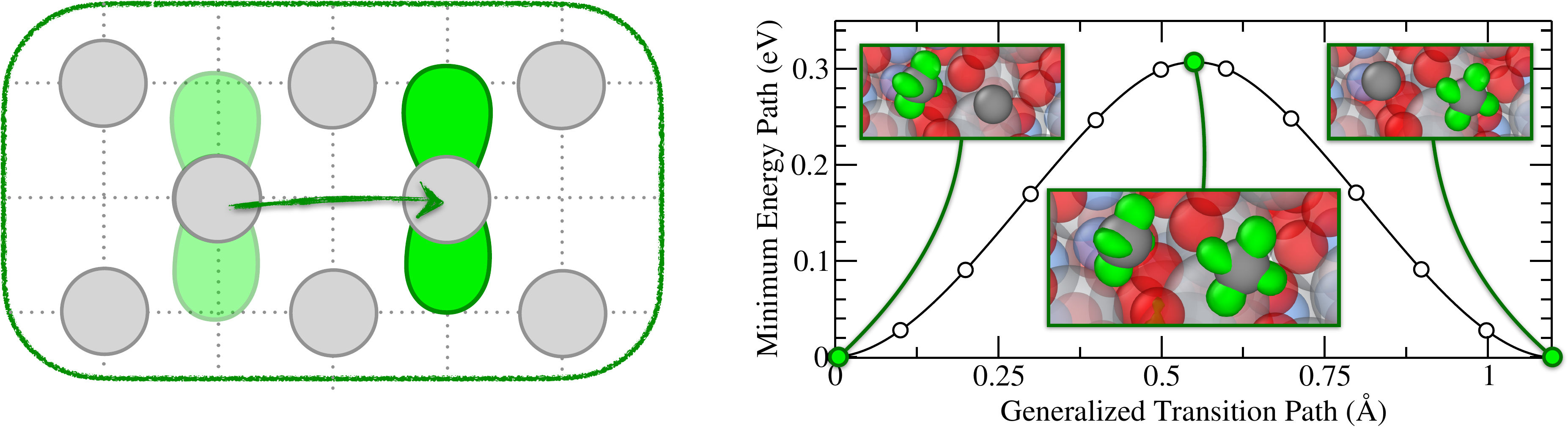}
    \caption{Polaronic transport at an atomistic level: While the left sketch exemplifies the underlying dynamics, the right plot shows the minimum-energy path for the hopping of a small electron-polaron between neighbouring Ti atoms in Li$_{4}$Ti$_{5}$O$_{12}$. The inlets showcase the initial, transition, and final state using the same color code of Figure~\ref{fig:polaron_lto}, whereby the involved Ti atoms are highlighted. At variance with {\it ionic} transport, cf.~Figure~\ref{fig:transport_ion},  only the electrons move and the nuclei stay on their lattice site in polaronic transport.}
    \label{fig:transport_pol}
\end{figure}
Eventually, polaronic conductivity constitutes a special case that encompasses ionic and electronic aspects. This is already evident from the temperature-dependence of the polaron viz. charge-carrier density. As the instructive example of anatase and rutile TiO$_2$ showcases~\cite{Setvin.2014,Franchini.2021}, the density of large polarons typically only features a weak temperature-dependence. Still, the respective conductivity typically decreases with temperature, suggesting that coherent band-type propagation as in the case of electronic transport is active. Accordingly, many-body perturbation theory as described above with appropriate extensions~\cite{Bernardi2019,chang2022intermediate} can be employed to model this electronic transport regime. In contrast, small polarons typically feature an activated behavior with both charge-carrier density and conductivity increasing with temperature. The latter reflects that small-polarons propagate through hopping, as in the case of ionic transport. However, charge and mass transport are not inherently linked and electronic degrees of freedom need to be explicitly taken into account, as showcased in Figure~\ref{fig:transport_pol}. Concurrently, the nuclear motion plays a formidable role, since the respective displacements are key to enabling a motion at all. Marcus theory~\cite{marcus1964chemical,fratini2016transient} can be invoked to describe this dynamics in terms of effective charge hopping events~\cite{oberhofer2017charge}. 
The probability for such events, the so called charge-transfer rates, can be derived using semi-classical transition state theory. The parameters entering this expression,~i.e.,~the effective vibrational frequencies and the Hamiltonian transition matrix elements, can be computed from first principles for each individual hopping process. The actual polaron dynamics can then be simulated by plugging these parameters in a kinetic Monte Carlo model; the transport coefficients are eventually extracted from the dynamics as described for ionic transport above.

\section{Challenges and Perspectives in Solar Battery Modelling}
\label{Chall}
\begin{figure}[t]
\centering
\includegraphics[width=.9\textwidth]{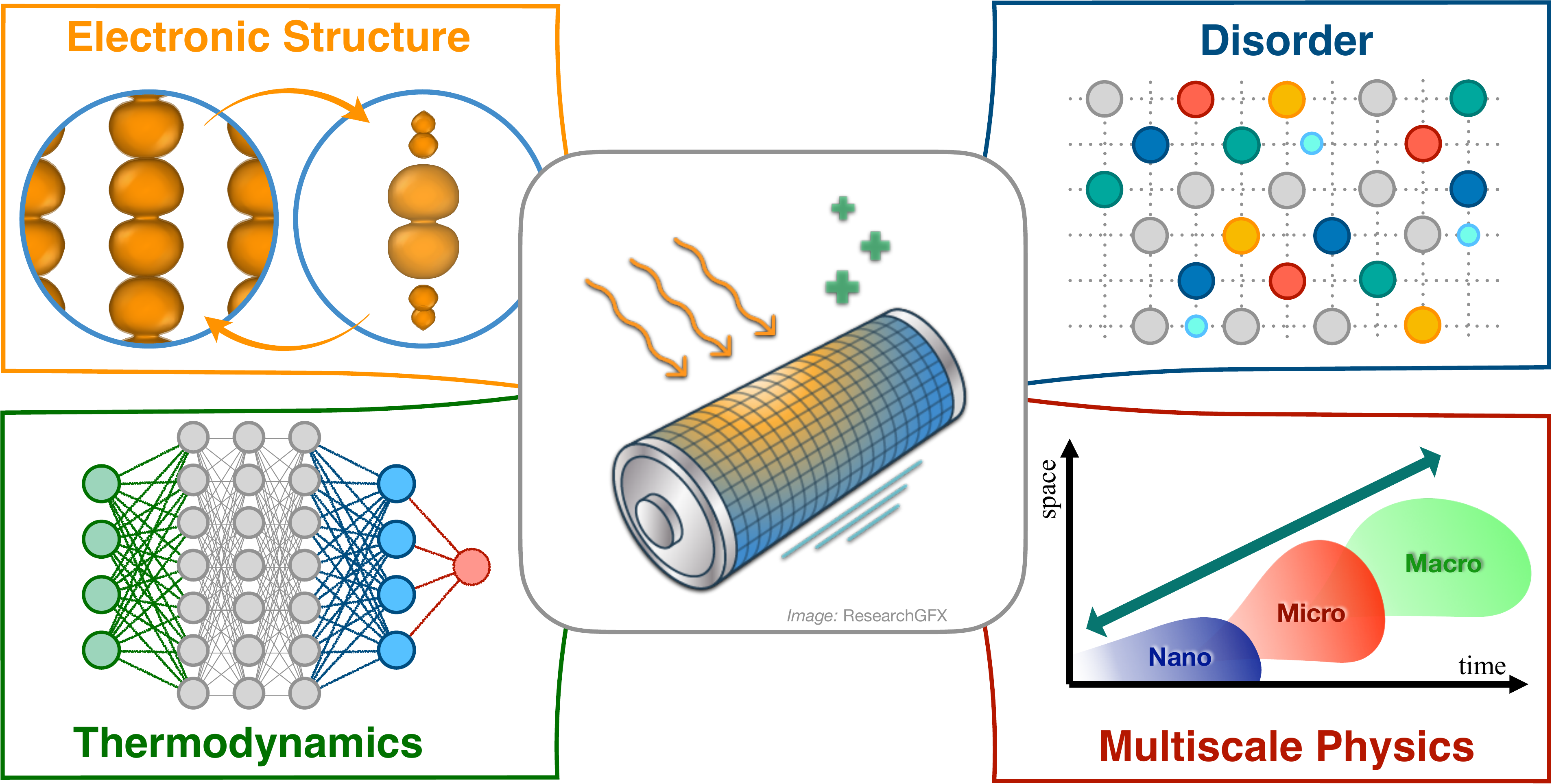}
\caption{Schematic overview of the challenges involved in the computational modeling of solar battery materials~(SoBaMs)}.
\label{fig:chall}
\end{figure}

As the previous section highlights, several computational approaches exist for tackling light-induced electron-hole pair creation, excess charge localization and stabilization, as well as charge transport. Nonetheless, modeling solar battery materials~(SoBaMs) is a formidable task, since the structural and stoichiometric richness of these compounds as well as the complexity of the involved mechanisms goes well beyond the present-day capabilities of the aforementioned methods. In the following, we discuss some of the specific conceptual and practical hurdles for SoBaM modeling, as well as prospects to overcome them in more detail.

\subsection{Methodological Complexity} 
As discussed in Sec.~\ref{sec:localization}, correctly describing charge localization is crucial for modeling SoBaMs. However, this typically requires electronic-structure theory beyond semi-local density-functional theory~(DFT). Semi-local exchange-correlation functionals suffer under the inherent self-interaction error~\cite{Cohen.2008}, resulting in artificially delocalized electrons and a severe underestimation of band gaps. Accordingly, charge localization and in particular polaron stability~\cite{Kokott.2018} are often severely underestimated in favor of more delocalized solutions both in ground state calculations and in excited-state approaches,~e.g.,~when modeling excitons~\cite{doi:10.1021/cr0505627,computation5010009}. 

Several {\it ad hoc} approaches have been proposed to circumvent this issue: For instance, the self-interaction can be removed analytically within many-body perturbation theory approaches, since one deals with an effective one-electron problem in this case, see (\ref{eq:sio_start}). In a similar spirit, different techniques to remove self-interaction by exploiting Janak's theorem have been developed for dealing with individual, isolated polarons~\cite{Sadigh.20157yo,falletta_manybody,Falletta.2022,Falletta.2023}. None of these approaches is easily applicable to the multi-polaron case of interest in SoBaMs, though.

More general self-interaction corrections have already been proposed in the very early days of DFT~\cite{Perdew.1981dqg} and are still topic of research today~\cite{Pederson.2014,Li.2017,Bi.2024,Schwalbe.2024}. One prominent and well-established technique to enforce localization at the semi-local DFT level with very modest computational overhead is the inclusion of Hubbard correction terms for localized electrons in $d$- or $f$-states, an approach commonly referred to as DFT+U~\cite{anisimov_1,lichtenstein,dudarev}. However, the strengths of these corrections are not known {\it a priori} and finding appropriate parameters can be challenging~\cite{QUA:QUA24521,PhysRevB.71.035105,Kick2019JCTC,PhysRevB.83.245124}. 

Another route to avoid self-interaction errors is to revert to a more advanced, albeit computationally more expensive electronic-structure theory method such as hybrid functional DFT~\cite{Cohen.2008}, which incorporates a fraction of exact Fock exchange. Although also in this case the amount of exact exchange that should be included to correct for self-interaction can be material-dependent~\cite{Atalla.2016,falletta_manybody}, the usage of hybrid functionals has been proven successful for modeling defects~\cite{batista2006comparison,agoston2009intrinsic,lyons2009nitrogen,clark2010intrinsic,komsa2011assessing} and polarons~\cite{Spreafico.2014,Janotti.2014,osterbacka2020small,wiktor2019electron,gono2018surface} since more than a decade. Naturally, even more advanced electronic-structure theory techniques,~e.g.,~the random-phase approximation~\cite{Kaltak.2014}, many-body perturbation theory~\cite{Chen.20158it}, quantum Monte Carlo algorithms~\cite{Parker.2011},  or coupled clusters approaches~\cite{Salihbegović.2023}, can be used as well and can provide even more accurate predictions for problematic, highly-correlated cases, if one can live with the considerably higher computational cost.

To conclude, correctly modeling localized electronic states and hence correcting for self-interaction effects is quintessential for modeling SoBaMs. {\it Ad hoc} corrections are generally preferable from a computational cost perspective, but are only applicable in certain cases,~e.g.,~for isolated polarons or when a +U correction is indeed sufficiently accurate. When this is not the case, hybrid-functional DFT calculations are the lowest rung of electronic-structure theory that can be sufficiently general and accurate, given that the electronic and physical mechanisms that are most relevant for charge localization in SoBaMs are incorporated from the start.

\subsection{Structural and Stoichiometric Complexity} 
Dealing with intercalation compounds, the modeling of SoBaMs typically requires to target structures featuring (several) hundreds of atoms or more to correctly capture structural and stoichiometric disorder, as already discussed in Sec.~\ref{sec:stabilization}.  As also mentioned in this context and further substantiated in the next subsection, machine-learned interatomic potentials are key to efficiently investigate the huge chemical space available for such materials. Before being able to use such MLIPs, it is however necessary to train them on sufficient {\it ab initio} data. This can constitute a severe practical and computational hurdle for SoBaMs, since this requires to run a large amount of advanced, beyond semi-local electronic-structure theory calculations for extended systems,~i.e., of at least several hundred atoms. For instance, the investigation of the SoBaM NbWO$_{6}$ system by Wang {\it et al.}~\cite{Wang.2024} was performed using electronic-structure theory calculations with hybrid functionals for 2D layers featuring 258 atoms so to ensure that localization of the photoelectrons is not artificially altered by finite-size effects. Although intercalation in this system is experimentally observed in solution, no explicit solvents were included in the theoretical modeling, since this would have increased the system size by at least another order of magnitude, making calculations largely unfeasible.

Already investigating the ground-state properties for system sizes with hundreds of atoms constitutes quite a challenge, since even semi-local DFT calculations scale cubically~$\sim\mathcal{O}(N^3)$ with system size~$N$ in standard implementations, since this is the scaling behavior of the computationally dominant Hermitian eigenproblem solver~\cite{Marek.2014,Kůs.2019}. Approaches with lower, even linear scaling~\cite{Hine.2009zpg,Bowler.2010,Bowler.2012} exist, but come with a substantial increase in prefactor with lower scaling. Especially for investigations featuring less than 10,000 atoms, it is hence not clear if lower scaling or increased prefactor dominates and a systematic benchmark of different approaches is advisable~\cite{Yu.2018}. Similarly, naïve implementations of the exact-exchange terms needed for hybrid DFT calculations feature a quartic scaling~$\sim\mathcal{O}(N^4)$, whereby typical implementations show a scaling between quadratic and cubic in practice~\cite{Dziedzic.2013}. Also in this case, exploiting localization viz.~the nearsightedness of electronic-structure theory~\cite{Kohn.1996,Prodan.2005} is key to establish linear-scaling algorithms~\cite{Gygi.2013}. In this regard, impressive progress has been made in recent years in increasing the computational efficiency and reducing the memory footprint of hybrid functional calculations, especially when it comes to large system sizes~\cite{Ko.2020,Ko.2023,Kokott.2024}. Exploiting these developments and further advancing these existing implementations holds great promise for addressing even the most complex SoBaMs at the required accuracy within tractable computational costs. 

Still, even these advancements can come to their limits, especially when it comes to modeling interfaces,~e.g.,~between electrodes and liquid electrolytes. In such cases, explicitly simulating every solvent molecule typically goes well beyond current computational capabilities. Accordingly, more approximate models need to be used for this purpose, for instance so-called implicit solvation methods, the usage of which is well established for modeling  electrochemical and electrocatalytic processes at solid-liquid interfaces~\cite{Ringe.2022}. These methods, which are based on a coarse-grained picture, do however not capture the full complexity of processes at the interface and hence need to be explicitly validated for the processes relevant for SoBaMs. 
Also in this context, the usage of MLIPs is a promising route towards establishing accurate and computationally manageable explicit solvation models. For this purpose, it is however necessary to develop MLIPs that at least qualitatively account for the essential electronic charge-transfer processes occurring in electrochemistry. More details in this regard are given in the next section.

An additional hurdle for modeling SoBaMs is, however, that targeting ground-state properties is not sufficient. Rather, excited state information is required as well, which poses an even more significant computational challenge, since such calculations are considerably more expensive than the respective ground-state computations. For instance, the most popular formalism is LR-TDDFT, which formally scales as~$\sim\mathcal{O}(N^6)$ due to the non-Hermitian eigenvalue problem at its core~\cite{10.1063/1.4919128}. In practice, a much lower scaling is typically achieved, given that one is typically not interested in the full excitation spectrum, but only in a few lowest lying excited states. This allows to leverage iterative eigensolvers such as the Lanczos \cite{doi:10.1063/1.2899649} or Davidson algorithm~\cite{DAVIDSON197587} and hence reduce the scaling of LR-TDDFT to $\mathcal{O}(N^3)$~\cite{doi:10.1021/acs.jctc.5b00887},  $\mathcal{O}(N^2)$~\cite{D0CP00060D}, or even $\mathcal{O}(N)$~\cite{doi:10.1021/ct200225v,10.1063/1.4936280,10.1063/1.2715568}. Also in this case, however, lower-scaling approaches come with a substantial increase in prefactor that can in practice still result in prohibitive runtimes.

To also reduce the prefactor, the community has developed strategies aiming at reducing the molecular orbital space, at truncating the necessary configuration space~\cite{D0CP06164F}, and/or at accelerating the evaluation of the necessary two-electron integrals involved in the eigenvalue equations,~e.g.,~by leveraging semi-empirical models~\cite{doi:10.1021/acs.jpca.1c02362,BANNWARTH201445}, density-fitting techniques with restricted auxiliary basis sets~\cite{10.1063/5.0020545}, or tight-binding approximations~\cite{Rger2016TightbindingAT,doi:10.1021/acs.jpcc.0c00979}. Especially when combined with each other~\cite{doi:10.1021/acs.jpca.1c02362,BANNWARTH201445}, these techniques have shown promise for addressing the low-lying excited states in large molecules and materials at a reasonable computational cost.

Although less popular than LR-TDDFT, RT-TDDFT has recently been gaining popularity especially for studying nonlinear effects in photoexcitation or various nonequilibrium electron dynamics phenomena.\cite{kanai_perspective,doi:10.1021/acsnano.0c03004,doi:10.1126/science.1249771,doi:10.1021/acsnano.8b08703,doi:10.1021/acsnanoscienceau.2c00061,doi:10.1021/acs.jpcc.7b04451,Rozzi_2018,doi:10.1126/science.1249771,https://doi.org/10.1002/qua.25096}. RT-TDDFT formally scales ~$\sim\mathcal{O}(N^2)$~\cite{O’Rourke.2015} where individual time steps in a RT-TDDFT typically have a computational cost that is comparable to that of a single ground-state iteration. Similarly, localization can be exploited in RT-TDDFT in the exact same fashion as in standard DFT to achieve linear scaling with system size~\cite{O’Rourke.2015}. However, the overall cost of a RT-TDDFT calculation can be substantial, given that the time-discretization typically requires time steps in the order of 0.0001-0.0005~fs. With that, applications of RT-TDDFT have been limited to rather short propagation times in the range of few femtoseconds or picoseconds at most.

Whenever these approximations hold, TDDFT efficiently simulates a few low-lying excited states in large molecules and materials. However, for systems requiring large supercells—such as nanosized materials in general and optoionic systems specifically—modeling becomes more complex due to defects, impurities, traps, or adsorbed species. The superlinear growth of surface excitons and mixing of bulk, surface, and molecular states lead to a high density of excited states. Since iterative methods compute one state at a time, this is inefficient when targeting higher-energy states or spectra, requiring many roots. Computational complexity can reach $O(m^3)$ for large $m$, with intermediate scaling of $O(m^2N^2)$ or $O(m^2N)$, making LR-TDDFT scale as $O(N^5)$ or $O(N^6)$ for large numbers of excited states. The high cost of LR-TDDFT and RT-TDDFT highlights the need for more efficient approaches to study systems at experimentally relevant sizes. For solar battery materials, this computational expense limits feasible system sizes to relatively small ones. However, this should still allow for the study of localized excitonic states in a somewhat idealized manner.

\subsection{Thermodynamic \& Statistical Sampling} 
As discussed in Sec.~\ref{sec:stabilization}, modeling intercalation compounds such as SoBaMs requires to sample a substantial phase-space volume, both in terms of chemical composition and of structural motifs. Exploring this humongous space --be it for calculating thermodynamic equilibrium or non-equilibrium properties-- requires millions if not billions of energy and forces evaluations, well beyond of what is reasonably possible at the {\it ab initio} level. As discussed in Sec.~\ref{sec:stabilization}, the usage of MLIPs provides a route to efficiently access long time and large length scales as well as the huge amount of combinatorial possibilities. Training such MLIPs does require accurate datasets, though, that faithfully cover the phase-space under exploration. While the usage of foundation models~\cite{takamoto2022towards,deng2023chgnet,merchant2023scaling,batatia2023foundation,zhang2024dpa,yang2024mattersim}
or active learning approaches that restrict data generation to minimal sets of maximum information content for the specific problem at hand~\cite{peng2022human,tran2018active,mou2023bridging,kunkel2021active,jinnouchi2020fly}
can help in this regard, more {\it ab initio} data for SoBaMs is in general urgently needed.

But even if enough first-principles data would be available, applying MLIPs for modeling SoBaMs is not as straightforward as one would hope. The fundamental reason is that the underlying optoionic effects are mediated by electrons and this electronic dynamics can often not be ignored and abstracted away, as usually done with MLIPs. In the simplest case,~i.e.,~in the case of small polarons, this results in atoms that are of the same chemical species, but feature a different oxidation state. By using geometric viz. chemical descriptors that reflect these different oxidation states of one and the same species, one can naturally incorporate this information into the MLIP during training. For instance, this enables studying polaron stability in configurational space, even in the presence of defects and other polarons~\cite{birschitzky2022machine}, or to build redox-aware MLIPs for investigating the oxidation-state patterns that correspond to the energetic ground state in battery materials~\cite{malica2024teaching}. However, this approach cannot be used to study polaron dynamics, since there is no explicit distinction between the electronic viz.~polaronic and the atomic~viz.~nuclear degrees of freedom. Accordingly, mass and charge transport are inherently linked as in ionic transport, see Figs.~\ref{fig:transport_ion} and~\ref{fig:transport_pol} and its discussions.

Various approaches have been proposed to (at least partially) incorporate and account for electronic degrees of freedom in MLIPs, starting from approaches that aim at predicting oxidation and spin states directly from local atomic environments~\cite{eckhoff2020predicting}. More general, charge-aware MLIPs with built-in coupling of electronic and nuclear degrees of freedom have also been developed. These efforts often
aim at improving the description of short-range interactions of local MLIPs by additionally including long-range electrostatic effects obtained by predicting local charges-states. Noteworthy developments include frameworks that rely on global charge equilibration~\cite{xie2020incorporating,ko2021fourth,vondrak2023q,rinaldi2024charge},
 approaches that incorporate total and/or local charge- and spin-constraints~\cite{unke2021spookynet,zubatyuk2021teaching,deng2023chgnet}, and several conceptually related methods~\cite{zhang2022deep,gao2022self,cools2022modeling} that encode the local charges by learning the locations of the maximally-localized Wannier centers~\cite{Marzari.2012}. In this spirit, it was recently also proposed to couple local magnetization predictions to the MLIP, so to incorporate polaron information in the architecture and to enable simulations of small-polaron dynamics~\cite{birschitzky2024machine_polaron}. 

While the above MLIPs formulated in terms of individual atoms and atomic environments are appropriate for modeling small polarons localized on specific atoms, they are hardly applicable to medium or large polarons, the size of which spans between several up to multiple unit cell. For this kind of systems, it has recently been proposed to accelerate existing many-body perturbation theory formalisms and to alleviate their numerical cost via data-driven techniques. For instance, it has been proposed to employ machine learning to predict electron-phonon coupling elements in reciprocal-space representation~\cite{li2024deep,zhong2024accelerating} or to compress them to lower dimensions, which also aids qualitative interpretation~\cite{luo2024data}. Still, these approaches are tightly bound both conceptionally and application-wise to the realm of many-body perturbation theory, cf.~Sec.~\ref{sec:localization}. More general and wider applicable techniques for learning not just energies and forces, but the whole electronic structure are still in their infancy, but hold great promise. For instance, this includes methodologies that leverage local real-space representations and then use equivariant methodologies to learn electron densities~\cite{Brockherde.2017,lewis2021learning} or Hamiltonians~\cite{zhang2022equivariant,li2022deep}.

The above concepts can be equally useful to accelerate excited-state simulations. Machine learning holds great potential to bypass the linear response equation, eliminating the expensive linear algebra computations in LR-TDDFT. More broadly, ML can map molecular structures not only to ground-state densities but also to excited states, enabling the prediction of excited-state potential energy surfaces and facilitating the study of the corresponding molecular dynamics. As a result, ML can accelerate calculations, allowing for the exploration of larger systems and more extensive conformational space searches. Indeed it has already proven to be particularly powerful in areas such as photochemistry, the description of excited states in molecules, and their dynamics.\cite{doi:10.1021/acs.jpclett.0c00527,doi:10.1021/acs.chemrev.0c00749,C9SC01742A,doi:10.1021/acscentsci.3c01480,D3CP05685F,Axelrod2022,Bai2022} However, applying machine learning to extended systems presents unique challenges, as the high density of excitations complicates the learning process. Additionally, accurately modeling long-range interactions to capture excitonic effects remains difficult for purely data-driven approaches.\cite{10.1063/1.1590951,Anstine2023} 
In terms of data efficiency, super-resolution techniques could serve as an excellent method. However, they have not been widely applied in the context of RT-TDDFT, as techniques such as compressed sensing typically require the dipole signal to consist of only a few well-separated frequencies—a condition often unmet in large systems. Recent advancements have shown that this requirement can be eased by introducing a physically motivated initial guess. First results for large systems are highly promising, indicating that a reduction in computational cost by a factor of 5 to 10 is more than realistic. This would ultimately allow for the precise extraction of excited-state information from short-time dynamics simulations. These findings could pave the way for a new super-resolution technique tailored specifically for excited-state simulations~\cite{Kick2024}.

To summarize, machine-learning techniques hold great promise for accelerating the development and advancing our understanding of SoBaMs. Still, this potential has not not yet been fully harvested. Although the electronic-structure methods discussed in the previous section constitute an excellent compromise between accuracy and efficiency, one common issue across all ML-based methods can still be the computational cost of acquiring sufficient and adequate training data. This can easily become computationally prohibitive, especially when it comes to extended systems, high degree of compositional disorder, or advanced methodologies such as TDDFT~\cite{10.1063/5.0047760}.  Especially for the latter case, transfer learning can offer a promising avenue to ensure high accuracy while improving data efficiency. The approach involves using lower-level theory to train the machine learning model and supplementing it with a limited number of data points from higher-level techniques to correct for missing physical effects. Similarly, integrating fundamental theoretical physics and chemistry knowledge into machine learning frameworks appears to be a promising route to improve on conventional data-driven models~\cite{Karniadakis2021,doi:10.1021/acscentsci.3c01480}. While these methods have proven effective for molecular systems, significant research is still needed to fully realize their potential for larger (extended) systems, as required for modeling SoBaMs. Along these lines, the above mentioned examples show that there has been a massive increase in interest in incorporating electronic degrees of freedom in MLIPs and in explicitly leveraging machine-learning techniques for directly predicting electronic properties, be it in the ground or in the excited state. These approaches hold huge potential for modeling SoBaMs, but have, so far not yet been applied to such complex materials. 

\subsection{Multi-Scale Feedback Loops}
So far, we have considered electron-hole pair dynamics, charge localization, and stabilization as separated processes that can be treated independently, also because the underlying physical mechanisms typically differ by orders of magnitude in time scales, see Sec.~\ref{sec:Intro}. This can be justified for modeling the thermodynamic equilibrium under the assumption that each of these processes has had enough time to equilibrate. However, such approximations are typically questionable when it comes to transport properties, for which the system is explicitly driven out-of-equilibrium.

Let us give just a few tangible examples for SoBaMs. For instance, ion intercalation can change the electronic-properties of a material, either directly 
or indirectly,~e.g.,~via lattice expansion. In turn, these changes can massively affect light absorption as well as electron-hole pair creation, recombination, and decay. Furthermore, it is known that hopping barriers for ionic conductivity can be lowered, but also increased by the nearby presence of (small) polarons~\cite{Kick.2021}. When simulating ionic conduction, it is hence crucial to account for this effect,~i.e.,~to account for the correlations between ionic and polaronic motion under the operating conditions of a SoBaM. Along these lines, it is very well possible that also the character of localized excess charge changes,~e.g.,~from a small polaron to a large one or vice versa, when the chemical environment changes~\cite{Verdi.2017}. At the meso- and macroscale, these effects can lead to the formation of a space-charge layer at interfaces~\cite{Komsa.2012,Richter.2013,Lehovec.1953,Bourelle.2024}, which is yet another possible mechanism underlying optoionic effects~\cite{Klotz.2020,Defferriere.2020,Defferriere.2022}. Accordingly, intercalation processes, concentration gradients in the material, or the interfaces between electrodes and electrolytes can alter the time scale of polaron transport by orders of magnitude, see Figure~\ref{fig:regimes} and its discussion.

Clearly, the listed problems call for the application of multiscale models that are able to bridge between several orders of magnitude in time and length scales. Such models are well established and explored in the realm of chemical reactions and catalysis and are hence readily available to model intercalation as well~\cite{Matera.2019,Bruix.2019}. However, much less is known about which formalisms and approximations are necessary and appropriate to incorporate electronic effects, including light-matter interactions~\cite{Hammerschmidt.2020} and the various flavors of charge transport. For the latter case, the massive advancements~\cite{Giustino.2017} made in the last decade in many-body perturbation can be regarded as a first step in establishing a unified picture, which in turn allows to treat electronic band-transport~\cite{Mustafa.2016,Poncé.2018}, small and large polaron formation~\cite{Lee.2021,lafuente2022_prb}, polaronic transport~\cite{Zhou.2019lc}, and couplings to excitons~\cite{Dai.2024jak,Dai.2024} on one footing. Still, extending these concepts to make them applicable to the complex, diverse, and heterogeneous materials of interest for SoBaMs is a formidable challenge. In this context, simplified, yet still accurate electronic-structure models such as density-functional tight binding hold great promise for reaching the necessary time and length scales~\cite{uratani2020quantum}.

\section{Conclusion}
\label{Concl}
This perspective aims at shedding light on the main physical and chemical mechanisms that are at the heart of the optoionic effect and of solar battery materials~(SoBaMs). These include electron-hole pair creation via above-band-gap illumination, charge separation and localization via defect and polaron formation, excess charge stabilization via intercalation, and various flavors of charge transport. For each of these aspects, we have concisely summarized the state-of-the-art theoretical and computational methodologies that are available today to address these mechanisms. Accordingly, we have discussed how to access excited electronic states, how to calculate the formation of localized excess charges, how advanced machine-learned interatomic potentials enable to tackle compositional and structural disorder, and how different transport regimes can be modeled. 

For modeling solar-battery materials, all of these aspects and methodologies need to be considered. This results in several computational and conceptual challenges that need to be tackled and that are outlined in this perspective. Two key hurdles arise from the fact that the optoionic effect is mediated by localized excess charges. This implies that the electronic-structure theory problem typically requires beyond semi-local treatments of exchange and correlation and that plain interatomic potentials that ignore electronic degrees of freedom are not applicable. Two further hurdles can be traced back to the coupling of electronic and ionic degrees of freedom. First, this results in a combinatorial explosion of possible chemical and structural modifications, since not only chemical species, but also different oxidation states, polarons, and charged defects need to be considered. Second, this can result in a coupling of electronic, polaronic, and ionic charge transport, which in turn requires the usage of multi-scale models to cover the various orders in magnitude in time and length scale that are involved. Computationally modeling these processes, understanding the underlying mechanisms, and the respective structure-stoichiometry-property relationships are key to advance this field. Ultimately, this would then pave the way for the accelerated discovery and design of SoBaMs via machine-learning-based inverse design, as recently already demonstrated for traditional batteries~\cite{liang2022review,ng2023machine,yu2023machine,wang2024materials}. 

Clearly, the hurdles on our path towards a computational modeling of SoBaMs are challenging. Recent advancements in electronic-structure theory, many-body perturbation theory, statistical mechanics, and artificial intelligence provide promising routes towards overcoming them, though. Most certainly, research in solar battery materials will be one the key driving forces in advancing this field in the years to come. In this context, let us emphasize that the described developments are not only relevant for solar-battery materials, but also more generally for modeling and understanding optoionic devices. For instance, this includes improving charge transport in conventional photovoltaics and electrolytes, controlling (potentially even dark) photocatalysis, and enabling photomemristive sensing as well as photoneuromorphic devices~\cite{Senocrate.2020,lv2022photoelectrochemical,Podjaski.2020}.

\section*{Data availability statement}
No new data were created or analyzed in this study.

\section*{Acknowledgements}
This work was generously supported by the Max Planck F\"orderstiftung with project SolBat. The authors acknowledge
inspiring discussions with Bettina V. Lotsch, Jennifer L. M. Rupp, and Christoph Scheurer.

\end{document}